\documentclass[a4paper]{jpconf}
\usepackage{graphicx}

\def\lsim{\raise0.3ex\hbox{$\;<$\kern-0.75em\raise-1.1ex\hbox{$\sim\;$}}}
\def\gsim{\raise0.3ex\hbox{$\;>$\kern-0.75em\raise-1.1ex\hbox{$\sim\;$}}}
\def\barr{\begin{eqnarray}}
\def\earr{\end{eqnarray}}
\def\beq{\begin{equation}}
\def\eeq{\end{equation}}

\newcommand{\ba}{\begin{array}{c}}
\newcommand{\ea}{\end{array}}

\def\nue{{\nu_e}}
\def\nuebar{{\bar{\nu}_e}}

\def\nux{{\nu_x}}

\def\pbar{\bar{p}}

\begin{document}
\title{Supernova neutrino oscillations: what do we understand?}

\author{Amol Dighe}

\address{Tata Institute of Fundamental Research, Homi Bhabha Road,
Colaba, Mumbai 400005, INDIA}

\ead{amol@theory.tifr.res.in}

\begin{abstract}

We summarize our current understanding of the neutrino flavor conversions
inside a core collapse supernova, clarifying the important role 
played by the ``collective effects'' in determining flavor conversion 
probabilities.
The potentially observable $\nu_e$ and $\bar{\nu}_e$ spectra   
may help us identify the neutrino mixing scenario, distinguish
between primary flux models, and learn more about 
the supernova explosion.

\end{abstract}

%%%%%%%%%%%%%%%%%%%%%%%%%%%%%%%%%%%%%%%%%%%%%%%%%%%%%%%%%%%%%%%%%%%
\section{Introduction}
%%%%%%%%%%%%%%%%%%%%%%%%%%%%%%%%%%%%%%%%%%%%%%%%%%%%%%%%%%%%%%%%%%%

Neutrino energy spectra exiting a core collapse supernova (SN)
are determined by the primary neutrino fluxes, the neutrino mixing
scenario as well as the densities encountered by the neutrinos 
along their path. (For recent overviews, see \cite{duan-review,nu08}.)
In this talk, we explain how the collective effects due to $\nu$-$\nu$ 
interactions and MSW matter effects result in distinctive features in 
the $\nu_e$ and $\bar{\nu}_e$ energy spectra at the detectors.
We emphasize some of the recent results involving the collective 
effects.

%%%%%%%%%%%%%%%%%%%%%%%%%%%%%%%%%%%%%%%%%%%%%%%%%%%%%%%%%%%%%%%%%%%%%
\section{Primary neutrino fluxes}
%%%%%%%%%%%%%%%%%%%%%%%%%%%%%%%%%%%%%%%%%%%%%%%%%%%%%%

Neutrinos from a SN are emitted in roughly three phases.
The first $\sim 10$ ms of the SN neutrino signal is 
the neutronization burst, composed entirely of $\nu_e$.
Neutrinos and antineutrinos of all species are emitted during
the next $\sim 10$ s, initially in the accretion phase and later
in the Kelvin-Helmholtz cooling phase \cite{book}.

The SN core acts essentially like a neutrino black body source
with flavor dependent fluxes. 
Since the primary fluxes of $\nu_\mu, \bar{\nu}_\mu, \nu_\tau, \bar{\nu}_\tau$
are almost identical, it is convenient to work in terms of the 
three ``flavors'':
$\nu_e \; , 
\nu_x \equiv \cos\theta_{23} \, \nu_\mu - \sin\theta_{23}\, \nu_\tau \; , 
\nu_y \equiv \sin\theta_{23}\, \nu_\mu + \cos\theta_{23} \, \nu_\tau$ ,
which are also the mass eigenstates at large matter densities. 
Clearly, the primary fluxes of $\nu_x, \nu_y, \bar{\nu}_x, \bar{\nu}_y$
are identical and we denote them as $F^0_{\nu_x}$.

The primary fluxes $F^0_{\nu_e}, F^0_{\bar{\nu}_e}, F^0_{\nu_x}$ are parameterized 
by the total number fluxes $\Phi_0$, average energies $E_0$, 
and the ``pinching parameters'' that characterize their spectral shapes 
\cite{keil1}.
The values of the parameters are highly model dependent, as can be seen from 
Table~\ref{tab:models}.
In the Table, M1 fluxes represent the predictions of the
Livermore model \cite{livermore}.
The predictions of the Garching model \cite{garching}, on the other hand,
are close to M1 during the accretion phase and close to M2 during
the cooling phase.
As will be seen later, the neutrino flavor conversion probabilities
depend on certain features of these primary spectra.

%%%%%%%%%%%%%%%%%%%%%%%%%%%%%%%%%%%%%%%%%%%%%%%%%%%%%%%%%%%%%%%%%%%%
\begin{table}[t]
\begin{center}
\caption{\label{tab:models}
Representative predictions of neutrino flux models.
$E_0$ values are given in MeV.}
\begin{tabular}{cccccc}
\br
{Flux Model} & 
${E_0(\nue)}$ &
${E_0(\nuebar) }$ &
${ E_0(\nux)}$ &
$\frac{\Phi_0(\nue)}{\Phi_0(\nux)}$ &
$\frac{\Phi_0(\nuebar)}{\Phi_0(\bar{\nu}_x)}$\\
\mr
{M1} & 12 & 15 & 24 &
 {2.0}&{1.6} \\
{M2} & 12 & 15 & 18 &
 {0.85 }& {0.75} \\
\br
\end{tabular}
\end{center}
\end{table}
%%%%%%%%%%%%%%%%%%%%%%%%%%%%%%%%%%%%%%%%%%%%%%%%%%%%%%%%%%%%%%%%%%%%%%

%%%%%%%%%%%%%%%%%%%%%%%%%%%%%%%%%%%%%%%%%%%%%%%%%%%%%%%%%%%%%%%
\section{Neutrino flavor conversions}
\label{conversions}
%%%%%%%%%%%%%%%%%%%%%%%%%%%%%%%%%%%%%%%%%%%%%%%%%%%%%%%%%%%%%%

%%%%%%%%%%%%%%%%%%%%%%%%%%%%%%%%%%%%%%%%%%%%%%%%%%%%%%%%%%%%%
\subsection{Collective effects at large neutrino densities}
\label{collective}
%%%%%%%%%%%%%%%%%%%%%%%%%%%%%%%%%%%%%%%%%%%%%%%%%%%%%%%%%%%%

The extremely high neutrino and antineutrino densities near the 
neutrinospheres make the $\nu$-$\nu$ interactions in this region significant
\cite{duan-fuller-carlson-qian-0606616,duan-fuller-carlson-qian-0608050}
and give rise to ``collective effects''.
Analytic studies of these effects reveal distinctive flavor 
conversion phenomena that are qualitatively different from the usual 
vacuum or MSW oscillations.
These include ``synchronized oscillations'' 
\cite{pastor-raffelt-semikoz-0109035} where
$\nu$ and $\bar{\nu}$ of all energies oscillate with the same frequency,
``bipolar/pendular oscillations'' 
\cite{hannestad-raffelt-sigl-wong-0608095,duan-fuller-carlson-qian-0703776}
that correspond to pairwise conversions
$\nu_e \bar{\nu}_e \leftrightarrow \nu_y \bar{\nu}_y$,
and ``spectral split'' 
\cite{raffelt-smirnov1,fuller-split,raffelt-smirnov2,nubar-split,multisplit}
where $\nu_e$ and $\nu_y$ ($\bar{\nu}_e$ and $\bar{\nu}_y$)   
spectra interchange completely within certain energy ranges.
The dynamics in three generations can be factorized into 
a superposition of multiple two-flavor phenomena \cite{threeflavor}.
New three flavor effects also emerge:
for example in early accretion phase, large $\mu$-$\tau$ matter potential 
may cause interference between MSW and collective effects
\cite{mutau-refraction}.

The dependence of the flavor evolution on the direction of
propagation of the neutrino may give rise to direction-dependent
evolution, or ``multi-angle effects'' 
\cite{duan-fuller-carlson-qian-0606616,duan-fuller-carlson-qian-0608050,Duan:2008eb,sawyer}
that may lead to decoherence
\cite{pantaleone-PRD58,raffelt-sigl-0701182,EstebanPretel:2008ni}.
However, for a realistic asymmetry between neutrino and antineutrino
fluxes, such multi-angle effects are likely to be small
\cite{fogli-lisi-marrone-mirizzi-0707.1998,estebanpretel-pastor-tomas-raffelt-sigl-0706.2498}
and the ``single-angle'' approximation  can be used.
Even for non-spherical geometries, one can study the single-angle evolution
along stream lines of neutrino flux, as long as 
coherence is maintained \cite{nonspherical}.

\begin{figure}
\includegraphics[width=20pc]{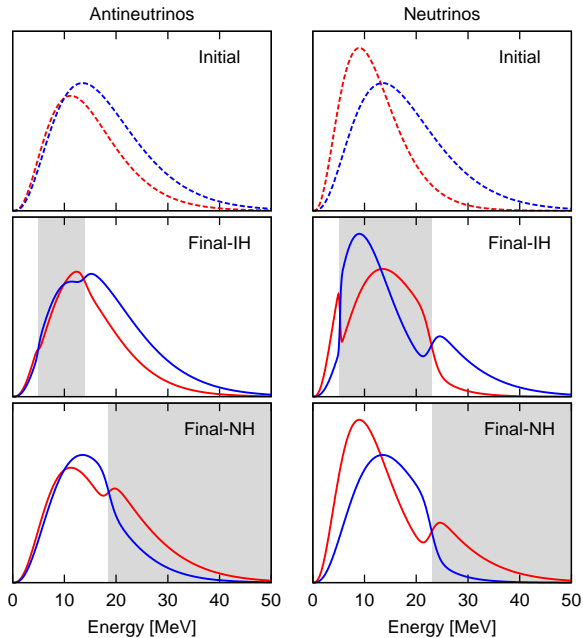}\hspace{2pc}%
\begin{minipage}[b]{12pc}\caption{\label{fig:multisplit}
The fluxes of antineutrinos and neutrinos (light grey/ red: $e$ flavor,
dark grey/ blue: $y$ flavor) before and after the action of
collective effects,
in the normal hierarchy (NH) and inverted hierarchy (IH). 
The shaded regions correspond to the swapped energy regime.
Here M2 fluxes from Table~\ref{tab:models} have been used.
The Figure has been adapted from \cite{multisplit}.}
\end{minipage}
\end{figure}

The propagation of the neutrinos can be rather cleanly separated 
into regions where various collective effects dominate individually
and hence the neutrinos experience these effects sequentially
\cite{threeflavor,fogli-lisi-marrone-mirizzi-0707.1998}.
When the neutrinos emerge from the region where collective effects
dominate, the net effect is a swap between the spectra of 
$\nu_e$ and $\nu_y$ ($\bar\nu_e$ and $\bar\nu_y$) in some energy
ranges, as illustrated in Fig.~\ref{fig:multisplit}.
The neutrino (antineutrino) spectra at this point can be divided into
two energy regimes: the ``swapped'' regime where the
$\nu_e$ and $\nu_y$ ($\bar\nu_e$ and $\bar\nu_y$) spectra have
completely interchanged, and the ``unswapped'' regime where the spectra have
remained unchanged.
At the boundary between two such regimes,
the survival probability changes sharply: this is the ``spectral split''.
This boundary need not be sharp: the spectral split is typically
smeared out over a small energy range.
In general, there can be zero, one, or multiple spectral splits in 
$\nu$ as well as $\bar{\nu}$ channels \cite{multisplit,irene}, 
and their positions can be approximately predicted depending on the 
mass ordering and primary spectra \cite{multisplit}.
Although the exact evolution of the neutrino state leading to the
splits can be calculated numerically (for animated simulations, see
{\tt http://www.mppmu.mpg.de/supernova/multisplits}),
the current analytic understanding of the dynamics is a bit unsatisfactory.
A clear direct mapping between the primary fluxes and 
split positions is also still lacking.

When the fluxes obey an approximate hierarchy,
i.e. $F^0_{\nu_e} > F^0_{\bar{\nu}_e} > F^0_{\nu_x}$ except
at high energies, the flux-split mapping is particularly straightforward.
All energies in $\nu$ as well as $\bar\nu$ are unswapped in normal
hierarchy, while in inverted hierarchy, only the energy range
$E < E_c$ for neutrinos below a certain critical energy $E_c$ remains 
unswapped \cite{raffelt-smirnov1,raffelt-smirnov2}.
The fluxes in the accretion phase (e.g. M1) typically fall under this
catagory.
The results in \cite{nu08,threeflavor,fogli-lisi-marrone-mirizzi-0707.1998,earth-hierarchy}, 
in particular, have been obtained with this flux scenario.

%%%%%%%%%%%%%%%%%%%%%%%%%%%%%%%%%%%%%%%%%%%%%%%%%%%%%%%%%%%%%%%%%%%%
\subsection{MSW resonances inside the SN and propagation through vacuum}
%%%%%%%%%%%%%%%%%%%%%%%%%%%%%%%%%%%%%%%%%%%%%%%%%%%%%%%%%%%%%%%%%%%%

In iron core supernovae, the collective effects have already become
insignificant when neutrinos enter the MSW resonance regions.
SN neutrinos must pass through two resonance layers: 
the H-resonance layer at
$\rho_{\rm H}\sim 10^3$ g/cc characterized by
$(\Delta m^2_{\rm atm}, \theta_{13})$,
and the L-resonance layer at
$\rho_{\rm L}\sim 10$ g/cc characterized by 
$(\Delta m^2_{\odot}, \theta_{12})$.
The outcoming incoherent mixture of vacuum mass eigenstates 
travels to the Earth without any further conversions,
and is observed at a detector as a combination of primary fluxes
of the three neutrino flavors:
$$
F_{\nue}  =  p F_{\nue}^0 + (1-p) F_{\nux}^0  \; , \quad \quad 
F_{\nuebar}  =   \pbar F_{\nuebar}^0 + (1-\pbar) 
F_{\nu_x}^0 ~,
$$
where $p$ and $\pbar$ are the
survival probabilities of $\nue$  and $\nuebar$ respectively.

The neutrino survival probabilities are governed by the adiabaticities 
of the resonances traversed, which are directly connected to the 
neutrino mixing scheme \cite{dighe-smirnov}.
Table~\ref{tab:pbar} shows the survival probabilities in swapped
and unswapped energy regions  in various mixing scenarios.
For intermediate values of $\theta_{13}$, i.e.
$10^{-5}\lsim\sin^2 \theta_{13} \lsim 10^{-3}$,
the survival probabilities depend on energy as well
as details of the SN density profile.
Note that though $p$ and $\pbar$ entries in scenarios C and D are identical,
since the swapped regime is determined by the mass hierarchy, even these
scenarios lead to different net flavor conversions.
The spectral swaps can occur for values of $\theta_{13}$
as low as $10^{-10}$ \cite{fuller-split,earth-hierarchy}. 
Collective effects are thus sensitive to extremely small (but nonzero) 
$\theta_{13}$, and can in principle allow identification of hierarchy.

%%%%%%%%%%%%%%%%%%%%%%%%%%%%%%%%%%%%%%%%%%%%%%%%%%%%%%%%%%%%%%%%%%%%%%%
\begin{table}[t]
\begin{center}
\caption{\label{tab:pbar} Survival probabilities $p$ and $\pbar$
in the swapped and unswapped energy regimes, 
in various mixing scenarios. Terms of ${\cal O}(\theta_{13}^2)$
have been neglected.}
\begin{tabular}{llccccc}
\br
{} &
{  Hierarchy} &  {$\sin^2\theta_{13}$}  & 
$p$ & $p$ & $\bar{p}$ & $\bar{p}$ \\
 & & & 
{unswapped} & {swapped} & {unswapped} & {swapped} \\
\mr
{  A} & Normal & {$\gsim 10^{-3}$}  & 0  & $\sin^2\theta_{12}$ 
& $\cos^2\theta_{12}$ & 0 \\
{  B} & Inverted &  {$\gsim 10^{-3}$} &  
$\sin^2\theta_{12}$ & 0 & 0 & $\cos^2 \theta_{12}$ \\
{  C} & Normal & {$\lsim 10^{-5}$}  & $\sin^2\theta_{12}$ & 0 
&  $\cos^2\theta_{12}$ & 0  \\
{  D} & Inverted & {$\lsim 10^{-5}$}  & $\sin^2\theta_{12} $ & 0
& $\cos^2\theta_{12} $ &   0 \\
\br
\end{tabular}
\end{center}
\end{table}
%%%%%%%%%%%%%%%%%%%%%%%%%%%%%%%%%%%%%%%%%%%%%%%%%%%%%%%%%%%%%%%%%%%%

%%%%%%%%%%%%%%%%%%%%%%%%%%%%%%%%%%%%%%%%%%%%%%%%%%%%%%%%%%%%%%%%%%%%
\subsection{Propagation through the shock wave near MSW resonances}
%%%%%%%%%%%%%%%%%%%%%%%%%%%%%%%%%%%%%%%%%%%%%%%%%%%%%%%%%%%%%%%%%%%

The passage of the shock wave through the H-resonance 
($\rho \sim 10^3$ g/cc) a few seconds after the core bounce may
break adiabaticity, thereby modifying the flavor evolution
of neutrinos that are emitted during this time interval
~\cite{fuller-shock,takahashi,ls1,lisi-shock,revshock,phase}.
Such a situation is possible, even in principle, only in certain
mixing scenarios: for neutrinos (antineutrinos), shock effects
can be present only in the mixing scenario A (B).
As a result, the mere identification of shock wave effects is enough 
to identify these scenarios, irrespective of the collective effects.

The shock wave effects can be diluted by stochastic density
fluctuations \cite{stochastic} or turbulence \cite{friedland}.
However, recent hydrodynamic simulations \cite{brockman,kneller}
suggest that some of the shock wave effects survive
in spite of these smearing factors.

%%%%%%%%%%%%%%%%%%%%%%%%%%%%%%%%%%%%%%%%%%%%%%%%%%%%%%%%%%
\subsection{Oscillations inside the Earth matter}
\label{earth}
%%%%%%%%%%%%%%%%%%%%%%%%%%%%%%%%%%%%%%%%%%%%%%%%%%%%%%%%%%

If neutrinos travel through the Earth before reaching the detector, 
they undergo flavor oscillations and the survival probabilities change
\cite{cairo,ls2,sato}.
This change however occurs only in the energy ranges where 
$p \neq 0$ (for neutrinos) or $\bar{p} \neq 0$ (for antineutrinos).
The presence or absence of Earth
effects in the neutrino or antineutrino channels would therefore
help in identifying some flux-mixing combination, using 
Table~\ref{tab:pbar}.

%%%%%%%%%%%%%%%%%%%%%%%%%%%%%%%%%%%%%%%%%%%%%%%%%%%%%%%%%%%%%%%%%
\section{Observable signals at neutrino detectors}
%%%%%%%%%%%%%%%%%%%%%%%%%%%%%%%%%%%%%%%%%%%%%%%%%%%%%%%%%%%%%%%%

If a SN explodes in our galaxy at $\sim 10$ kpc from the Earth, 
we expect ${\cal O}(10^4)$ events of $\bar{\nu}_e$ at Super-Kamiokande (SK). 
With future larger water Cherenkov detectors, large scintillation detectors 
with better energy resolutions, and liquid Argon detectors sensitive 
to $\nu_e$, it should be possible to reconstruct the $\nu_e$ and
$\bar{\nu}_e$ spectra to a good accuracy. 
SN neutrinos would also enable pointing at the SN in advance of the 
optical signal \cite{snews}, with \cite{kate-pointing} or even without 
\cite{beacom,ando,pointing} any information on neutrino mixing,
and reconstructing the SN bounce time accurately
\cite{raffelt-halzen}.

%%%%%%%%%%%%%%%%%%%%%%%%%%%%%%%%%%%%%%%%%%%%%%%%%%%%%%%%%%%%%%
\subsection{Suppression of $\nue$ in the neutronization burst}
%%%%%%%%%%%%%%%%%%%%%%%%%%%%%%%%%%%%%%%%%%%%%%%%%%%%%%%%%%%%%

Since the primary signal during the neutronization burst is pure $\nu_e$,
the entire energy range is unswapped.
Since the model predictions for the energy and luminosity
of the burst are fairly robust \cite{ricard-neutronization},
the observation of the burst signal gives direct information
about the survival probability of $\nu_e$. This probability is
${\cal O}(\theta_{13}^2)$ in scenario A and $\sin^2 \theta_{12}$
in all the other scenarios \cite{dighe-smirnov}. 
Thus, the strong suppression of $\nu_e$ burst would be a smoking gun signal 
for scenario A.

In an O-Ne-Mg supernova, the MSW resonances may lie deep inside
the collective regions during the neutronization burst, when
the neutrino luminosity is even higher.
In such a situation, neutrinos of all energies undergo the
MSW resonances together, with the same adiabaticity \cite{wong}. As long
as this adiabaticity is nontrivial, one gets the ``MSW-prepared
spectral splits'', two for normal hierarchy and one for inverted
hierarchy \cite{fuller-split1,fuller-split2,onemg-doublesplit}.
The positions of the splits can be
predicted from the primary spectra \cite{onemg-doublesplit}.
The splits imply $\nu_e$ suppression that is stepwise in energy.
Such a signature may even be used to identify the O-Ne-Mg
supernova, in addition to identifying the hierarchy.

%%%%%%%%%%%%%%%%%%%%%%%%%%%%%%%%%%%%%%%%%%%%%%%%%%%%%%%%%%%%%%%%%%%
\subsection{Shock wave effects}
\label{shock}
%%%%%%%%%%%%%%%%%%%%%%%%%%%%%%%%%%%%%%%%%%%%%%%%%%%%%%%%%%%%%% 

Observables like the number of events, average energy,
or the width of the spectrum may display dips or peaks 
for short time intervals, while the shock wave is
passing through the H resonance.
The positions of the dips or peaks in the number of events at different 
neutrino energies would also allow one to trace the shock
propagation while the shock is in the mantle, around densities of
$\rho \sim 10^3$ g/cc \cite{revshock}.
This information, by itself or in combination with the corresponding
gravitational wave signal, will yield valuable information about 
the SN explosion, in addition to identifying the scenario A or B.

If light sterile neutrinos exist, they 
may leave their imprints in the shock wave
\cite{choubey-ross1,choubey-ross2}.
For an O-Ne-Mg supernova, passage of the shock wave through
the sharp density profile at the resonance leads to
distinctive effects \cite{lunardini-onemg}.

%%%%%%%%%%%%%%%%%%%%%%%%%%%%%%%%%%%%%%%%%%%%%%%%%%%%%%%%%%%%%%%%
\subsection{Spectral split in $\nu_e$}
%%%%%%%%%%%%%%%%%%%%%%%%%%%%%%%%%%%%%%%%%%%%%%%%%%%%%%%%%%%%%%%
                                              
A split in $\nu_e$ or $\bar{\nu}_e$ spectrum would manifest itself 
as a sharp jump at the split energy \cite{fuller-split}. 
It is therefore a smoking gun signal for collective effects
and the corresponding flux-mixing combination.
However, the sharp split in $p$ is smeared to some extent by the
multi-angle effects 
\cite{fogli-lisi-marrone-mirizzi-0707.1998}
and its actual observation may be a challenging task.

%%%%%%%%%%%%%%%%%%%%%%%%%%%%%%%%%%%%%%%%%%%%%%%%%%%%%%%%%%%%%%%
\subsection{Earth matter effects}
%%%%%%%%%%%%%%%%%%%%%%%%%%%%%%%%%%%%%%%%%%%%%%%%%%%%%%%%%%%%%%

Earth matter effects can be identified by the comparison of signals 
at two detectors, only one of which is shadowed by the earth.
This could be achieved through the $\bar\nu_e$ spectra 
at two large water Cherenkov detectors \cite{earth-hierarchy}
or through the time dependent ratio of luminosities
at IceCube and Hyper-Kamiokande \cite{ice-hyper}. 
The Earth effects can be identified even at a single detector
as long as it is capable of determining the neutrino energy.
For example, Fourier transform of the ``inverse energy'' spectrum of 
$\nuebar$ \cite{fourier}
gives peaks (multiple ones if the neutrinos traverse the Earth core)
whose positions are independent of the primary neutrino 
spectra. These would reveal the presence of Earth matter effects
\cite{corewiggles}.
For the typical accretion phase scenario (M1) and
$\sin^2 \theta_{13} \lsim 10^{-5}$, the detection of Earth effects 
in $\bar\nu_e$ can identify normal hierarchy \cite{earth-hierarchy}.

The presence of spectral splits implies that the Earth effects 
may be present in only certain energy regimes. The identification
of these energy regimes can play a major role in the identification
of the corresponding flux-mixing combination.

%%%%%%%%%%%%%%%%%%%%%%%%%%%%%%%%%%%%%%%%%%%%%%%%%%%%%%%%%%%%%%%%%
\section{Concluding remarks}
%%%%%%%%%%%%%%%%%%%%%%%%%%%%%%%%%%%%%%%%%%%%%%%%%%%%%%%%%%%%%%%%%

The flavor conversion probabilities of neutrinos inside a SN 
are sensitive to the neutrino mixing scenarios, in particular to 
the mass hierarchy even at extemely small $\theta_{13}$,
thanks to the collective effects and MSW matter effects.
In addition, since the collective effects can change the neutrino flavor
composition deep inside the core, they may affect the dynamics of
the SN explosion.

Smoking gun signals of neutrino mixing scenarios as well as information 
on primary fluxes and shock wave propagation can be independently 
obtained through observations like the suppression of neutronization burst, 
time variation of the signal during shock wave propagation, 
and Earth matter effects.
More reliable predictions of primary fluxes, better understanding of 
the collective effects, and improved limits on $\theta_{13}$ from
terrestrial experiments will make our deductions more robust.

\ack

I would like to thank Basudeb Dasgupta, Alessandro Mirizzi,
Georg Raffelt, and Alexei Smirnov for fruitful collaborations, 
the organizers of
TAUP 2009 for their hospitality, and the Max Planck - India
Partnergroup for financial support.

\section*{References}
%%%%%%%%%%%%%%%%%%%%%%%%%%%%%%%%%%%%%%%%%%%

\end{document}